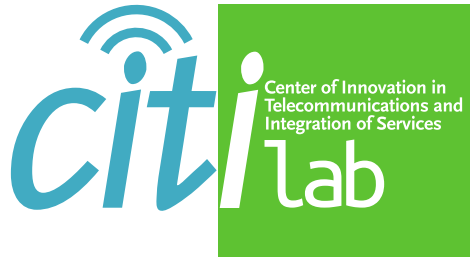

SCIENTIFIC REPORT

# Usability and Performance Analysis of Embedded Development Environment for On-device Learning

Enzo Scaffi [1]   Antoine Bonneau [1,2]   Frédéric Le Mouël [1]

Fabien Mieyeville [2]

[1] INSA Lyon, Inria, CITI, UR3720, 69621 Villeurbanne, France
{antoine.bonneau, frederic.le-mouel}@insa-lyon.fr

[2] Université Claude Bernard Lyon 1, INSA Lyon, Ecole Centrale de Lyon, CNRS, Ampère, UMR5005, 69622 Villeurbanne, France
fabien.mieyeville@univ-lyon1.fr

March 14, 2024

**ABSTRACT**    This research empirically examines embedded development tools viable for on-device TinyML implementation. The research evaluates various development tools with various abstraction levels on resource-constrained IoT devices, from basic hardware manipulation to deployment of minimalistic ML training. The analysis encompasses memory usage, energy consumption, and performance metrics during model training and inference and usability of the different solutions. Arduino Framework offers ease of implementation but with increased energy consumption compared to the native option, while RIOT OS exhibits efficient energy consumption despite higher memory utilization with equivalent ease of use. The absence of certain critical functionalities like DVFS directly integrated into the OS highlights limitations for fine hardware control.

# I  INTRODUCTION

This research aligns with developing a small wireless device intended to function within networks of similar devices, operating independently of external power and communication networks. Within this scope, we are contemplating the software development for an embedded system, an autonomous computing device integrated into an object or a more extensive system capable of performing specific tasks. It is less potent than a traditional computer but significantly more energy-efficient and more in scope with ubiquitous computing. For instance, we might equip this device with a battery and energy-harvesting solution such as a small solar panel for autonomy. The idea is to create a system roughly the size of the palm of a hand or even smaller. We aim to design an intelligent sensor capable of processing data and making decisions, categorized into two phases: learning and inference using ML algorithms.

Pursuing this objective, we explore key concepts crucial to our embedded system's development within a wireless edge network. These concepts include intermittent learning [1], [2], which adapts learning processes to conserve energy, aligning with the system's low-energy consumption goals. Our system aims to embody ambient intelligence [1] by reacting to environmental changes, resonating with the IoT principle [3] as part of an interconnected system that collects, processes, and transmits data. We consider Federated Learning's collaborative approach, aiming to improve performance without centralizing data. We fall into Frugal AI since we will need our resource-efficient AI to align with our system's need to execute complex tasks with minimal energy consumption and scarce data availability. On-device learning [2], [4] and TinyML [5], [6] play integral roles in acquiring knowledge independently and implementing advanced on-device ML techniques, fitting perfectly within our embedded system's operational environment.

In Section II, we present general constrained device classification and associated ML tasks. We present our hardware targets in Section III and the tools selected for this study in Section IV. Section V outlines the measurements conducted and the evaluated applications. Finally, in Section VI, we compile the results from various experimental measures.

# II  EMBEDDED ML DEVICES CLASSIFICATION

Applying Machine Learning (ML) in embedded systems is not new. Edge computing represents an innovative approach to bring data processing closer to its sources (i.e., sensors, cameras, etc.). This approach leverages interconnected smart devices for decentralized computation. Edge computing presents critical implications for the Internet of Things (IoT), including latency reduction, privacy preservation, failure resilience, and system modularity. [7]

The literature proposes different classifications of constrained devices used in edge processing. This classification from Bonneau et al. [8] is based on the hardware character-istics, and we added energy consumption and ML algorithms available. Distinct device categories are identified in Table 1, ranging from Class 0 to Class 19. Class 0 to Class 4 encompasses microcontrollers with limited memory and power, suitable for lightweight inference. Class 10 to Class 19 devices gather high-performance application processors with substantial memory and power and fall out of the scope of this study. Those devices are usually selected for complex tasks such as deep learning, training, and advanced inference. Each device class is associated with specific ML algorithms based on their capabilities, ranging from simplicity to complexity.

| Class Name | Memory Size | | Average Power Consumption | Typical Device | ML Use Case |
|---|---|---|---|---|---|
| | Data e.g. RAM | Code e.g. Flash | | | |
| Class 0 C0 | $\ll$ 10 KiB | $\ll$ 100 KiB | $\leq$ 100 µW | RFID Tag | Basic computations (lightweight inference) |
| Class 1 C1 | $\simeq$ 10 KiB | $\simeq$ 100 KiB | $\leq$ 1 mW | Basic Sensor | |
| Class 2 C2 | $\simeq$ 50 KiB | $\simeq$ 250 KiB | $\leq$ 10 mW | IIoT Sensors | |
| Class 3 C3 | $\simeq$ 100 KiB | $\simeq$ 0.5..1 MiB | $\leq$ 100 mW | Wearable IoT | Basic statistics (inference) |
| Class 4 C4 | $\simeq$ 0.3..1 MiB | $\simeq$ 1..2 MiB | $\leq$ 1 W | IoT Gateway | |
| Class 10 C10 | $\simeq$ 16..128 MiB | $\simeq$ 4..16 MiB | $\leq$ 2 W | OpenWRT routers | Classification, Regression (inference) |
| Class 15 C15 | $\simeq$ 0.5..1 GiB | $\simeq$ 16..64 MiB | $\leq$ 4 W | Raspberry PI | |
| Class 16 C16 | $\simeq$ 1..4 GiB | *(lots)* | $\leq$ 16 W | Smartphone | Prediction Decision-making (inference & training) |
| Class 17 C17 | $\simeq$ 4..32 GiB | *(lots)* | $\leq$ 100 W | Laptop | |
| Class 19 C19 | *(lots)* | *(lots)* | $\gtrsim$ 1000 W | Server | Deep learning, auto-encoders (inference & training) |

**Table 1.** Classes of constrained devices associated with ML use cases

## III  BOARDS SELECTION

Minimizing energy consumption involves employing low-power (LP) [7] or even ultra-low-power (ULP) boards. Several criteria classify embedded electronic boards as LP or ULP [8], [9]. The first criterion considers energy consumption during board inactivity, while the second focuses on energy usage during activity. Additionally, standby power consumption is a critical factor. Examining the components constituting the board for energy-saving designs is essential. Evaluating power management methods, such as dynamic power management and contention, which adjust power based on workload and battery, is also crucial [10]. We assessed several boards based on various aspects:

- **Performance**: Not all boards have the same computational or memory capabilities, necessitating performance testing.
- **Energy Consumption**: This is a primary low-consumption consideration. Conducting comprehensive measurements in different scenarios (and implementations) is essential to understand the board's energy behaviour beyond what datasheets offer.
- **Functionalities**: Boards vary in features; some integrate telecommunication systems while others don't. Certain boards support specific frameworks.
- **Development Ease for Hardware Target**: This pragmatic aspect is crucial. Not all boards embed the same basic software component and are powered by fifferent community more or less active and sharing specialized tools; some present more development challenges than others.

Several board families met the previously stated criteria in Bonneau et al.'s work [8]. However, we had to work on different versions of the nodes curated due to time limits and hardware availability. We worked with the STM32 L073RZ from ST Microelectronics and a first-generation generic ESP32 from Espressif (Table 2). Those two boards are nevertheless representative of computing and memory capacity of targetted boards. The objective was to test program implementation on these platforms. However, it is essential to note that we did not conduct energy consumption tests on the ESP32 due to its high consumption at boot, which caused the measurement board to fault. Nevertheless, given the code's portability, newer variants free from overvoltage issues can execute the provided code.

| IoT Board | Processing Unit | Clock Speed (MHz) | Memory (kB) | On-board Storage (kB) | Electrical Characteristics | Wireless Connectivity | Release & Price* |
|---|---|---|---|---|---|---|---|
| | **A**pplication Core | | **N**etwork Core | | **R**egulated | **U**nregulated | |
| STM32 L073RZ | Arm Cortex M0+ 32-bit RISC | 64 | 20 (RAM) 6 (ROM) | 1024 | **R** 1.8 - 3.6 V | — | 2015 13 € |
| ESP32 DK M1 | Xtensa LX6 dual-core 32-bit | 80 to 240 | 520 (RAM) 448 (ROM) | 4096 | **R** 3.3 ± 0.3 V **U** 5..12 V | WiFi Bluetooth | 2020 8 € |

**Table 2.** Selected devices among radio-enabled Middle-end IoT boards

# IV  EMBEDDED DEVELOPMENT AND RUNTIME ENVIRONMENT

The goal is to evaluate the overhead energy use of different applications from different abstraction levels. We consider three development methods: coding natively for the ST board using STM32CubeIDE tools, Arduino Framework through PlatformIO, and RIOT OS for IoT devices. We want to evaluate how each method varies in performance, memory usage, and energy efficiency for embedded systems, offering trade-offs in energy consumption, portability, and task management.

### 1. Native Approach

Developing directly on hardware without intermediary operating systems or software layers, known as the *native* or *bare metal* approach, implemented using C++ code, grants complete control over resources, enabling energy consumption optimization by turning off unnecessary hardware components and fine-tuning code execution. This method allows for adjusting energy usage at its lowest level. However, its complexity demands extensive hardware knowledge, making development intricate, while its lack of portability restricts applications to specific hardware, complicating transitions to other platforms.

### 2. Arduino Framework Approach

The Arduino Framework approach simplifies embedded app development through a user-friendly interface featuring pre-built libraries, expediting project creation and enabling C-like code reuse across platforms. It fosters portability by allowing applications to transfer between compatible Arduino boards without extensive code rewriting. However, while offering ease of use with simpler code and ready-to-use libraries, this approach might incur higher energy consumption due to the abstraction it introduces, potentially leading to increased overhead compared to a native approach.

### 3. Operating System-based Approach

Leveraging an operating system such as RIOT OS, written in C, C++, provides task management, inter-task communication, and memory control, enabling the development of intricate applications and effectively handling simultaneous resources. Operating systems offer heightened portability across diverse hardware platforms through hardware abstraction layers, facilitating seamless migration. They also support concurrent execution of multiple tasks, streamlining development with precise syntax and pre-built libraries, employing a high-level language. However, this approach may introduce increased energy consumption due to inherent overheads, and configuring the system for low-level programming without direct APIs can pose complexities.

Testing the three approaches (Native, Arduino Framework, and OS) in similar use cases allows for precise evaluation of energy consumption costs and programming feasibility using these tools, while considering the specificities of each hardware platform. In reality, each approach strikes a balance between advantages and drawbacks, notably regarding energy consumption. However, in this project, multitasking capability is essential, and we aim to assess the additional energy impact of employing an operating system. Nevertheless, the actual impact on energy consumption can vary significantly based on factors such as application complexity, executed operations, and code optimization.

### IV.B  Discussion on the Choice of the Operating System

Using an OS provides access to additional elements compared to a framework. Firstly, file management which involves a directory structure containing other folders or files, similar to what we use on our computers daily. In our case, while this could store our model when we have no battery, it could already be achieved by saving context in flash memory. Secondly, multitasking is another advantage an OS brings, allowing us to execute multiple tasks simultaneously. In practice, this involves scheduling processor instructions by alternating between different tasks and it could enable continual learning while using the communication module to send data to the user or communicate with other systems.

Bonneau et al. [8] proposed several real-time operating systems (RTOS) that are usable within the scope of machine learning on communicating nodes listed in Table 3. RIOT OS stands out, boasting compelling APIs for energy management, enabling native access to alternative operating modes of equipped microcontrollers. These modes facilitate selective standby of hardware components when not in use, effectively reducing energy consumption during inactive periods. It also offers a highly active community, excellent documentation, and compatibility with numerous consumer-grade boards.

| RTOS | Platforms | RAM (kB) | ROM (kB) | Codebase (MLOC) | Programming Language | Scheduling | Last Release |
|---|---|---|---|---|---|---|---|
| Apache MyNewt | ARM (M0/3/4/23/33) MIPS32, Microchip PIC32, RISC-V | ~ 1 | ~ 10 | ~ 1.2 | C, C++ | Preemptive Priority based | 1.11.0 Sept. 2023 |
| Apache Nuttx | ARM, AVR8, MIPS, Renesas, RISC-V, Xtensa, ZiLOG | – | – | ~ 4.3 | C, C++ | Preemptive Priority based | 12.2.1 Jul. 2023 |
| Zephyr | ARM, x86, ARC, RISC-V, Nios II, Xtensa, SPARC | – | ~ 2 to ~ 8 | ~ 3 | C, C++ | Preemptive, Priority based, Cooperative | 3.4.1 Jun. 2023 |
| FreeRTOS | ARM, AVR(32), ColdFire, Xtensa, HCS12, RISC-V, IA-32, Infineon XMC4000, MicroBlaze, MSP430, PIC(32), Renesas H8/S, RX100/200/600/700, 8052, TriCore, EFM32 | ~ 1 | ~ 5 to ~ 10 | ~ 7.2 | C, C++, Go (Rust Wrappers) | Preemptive, Priority based, Cooperative | 10.6.1 Aug. 2023 |
| RIOT | ARM, MSP430 AVR, x86, RISC-V | ~ 1.5 | ~ 5 | ~ 3 | C, C++ (Rust Wrappers) | Preemptive, Priority based | 2023.07 Aug. 2023 |

**Table 3.** Low-End and Middle-End IoT OSs comparative overview (from [8])

## V EXPERIMENTS AND METHODOLOGY

We have conducted a series of experiments to evaluate the performance and trade-offs of the different approaches. We wanted to implement each curated application with those approaches but faced some difficulties as mentioned later in Section VI.

### V.A SURVEYED APPLICATIONS

We have selected three applications that involve different levels of complexity and correlation with hardware knowledge:

- **Blinky:** A straightforward program that simply turns on the integrated LED on the MCU-equipped board (most boards have at least one controllable LED). The LED lights up for 1 second, then turns off for 1 second, looping continuously.
- **Frequency:** A program that gradually decreases the MCU frequency, starting from the highest available frequency and progressively reducing it until reaching the lowest available frequency on the main power mode.
- **Classifier:** A program that conducts basic ML using a perceptron model [11]. We train on a portion of each dataset and test on another. The datasets used are well-known and easily implementable, such as Iris flower [12], Heart Disease [13], Breast Cancer [14], MNIST Handwritten Digits [15]. We have reduced the larger ones to fit our boards' memory constraints.

## V.B Instrumentation and Measurements

We utilized the X-NUCLEO-LPM01A [16], a programmable power supply with a voltage range from 1.8 V to 3.3 V, enabling static (from 1 nA to 200 mA) and dynamic (from 100 nA to 50 mA) current measurements with up to 100 kHz of sampling frequency. Leveraging the software STM32CubeMonitor-Power [17], an analysis of energy consumption for target boards was conducted. This software facilitates the acquisition of power measurements and displays them through a graphical interface. It is not recommended to perform acquisitions superior to one hour at max rate or over 360 million samplings, but only due to GUI issues. Furthermore, it enables storing and retrieving acquired data saved in CSV files. Each version of the application was executed on every platform within predefined scenarios, capturing precise data regarding energy consumption for each run.

Comparison was made among the energy consumption data acquired for each approach and platform, identifying significant differences among approaches. Subsequently, all variability factors potentially influencing the outcomes were considered, including hardware discrepancies (RIOT, Arduino Framework, and ST SDK do not activate the same functionalities by default on the same board; these activated or deactivated functionalities impact efficiency, and sometimes optimizations are made on specific platforms) and the quality of implementation (code differs for each platform).

## VI EXPERIMENTAL RESULTS

The following programs were developed on the STM32 L073RZ platform, operating at a frequency of 32 MHz. Data is extracted and analyzed using a Python script. This script handles the processing of current measurements, energy consumption information, and memory usage details. The obtained data is then transformed into graphs and visual representations for more in-depth analysis and a better understanding of the performance and characteristics of each test.

## VI.A Memory Usage Analysis

Our focus was initially on memory usage during each test, and these data were recorded in a JSON-formatted document. This approach allows us to have a precise view of the memory footprint of each test and to compare this characteristic across different configurations.

### 1. Blinky

We can observe in Figure 1 that using a framework such as Arduino (PlatformIO) or RIOT indeed results in a greater need to store data in Flash memory, as expected. The difference between Arduino Framework and RIOT concerning ROM memory usage remains relatively small (less than 5%), which is excellent news. The use of ROM memory for this application remains moderate, as even with RIOT, we do not exceed 10% usage. The RAM analysis is not particularly significant, considering we use very little volatile memory to blink our LED (assuming our board's maximum capacity is 20480 bytes). However, a trend seems to emerge: RIOT consumes slightly more RAM, which is expected due to its more advanced features.

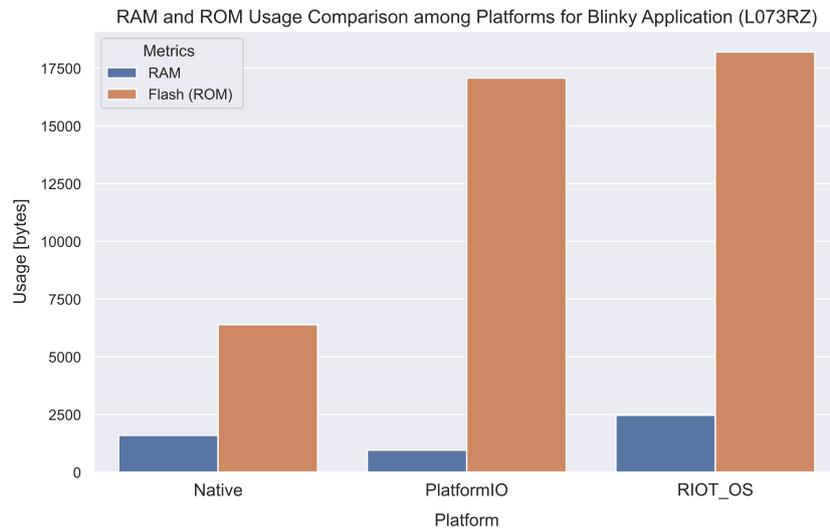

**Figure 1.** Comparison of RAM and ROM Usage Across Platforms for Blinky

## 2. FREQUENCY

Figure 2 presents significant results. A decrease of approximately 40% in RAM usage is observed when using PlatformIO compared to the original STM32 tool. In both cases, RAM usage remains relatively low. However, ROM memory usage shows a notable increase of over 70% when using PlatformIO compared to the ST tools. Ultimately, we find ourselves in a similar scenario to the previous experiment; this is normal as we use functions that do not require much memory. It is important to note that this experiment demonstrates that both tools manage memory differently, which is expected given their different approaches. This observation emphasizes that the tools handle memory in different ways.

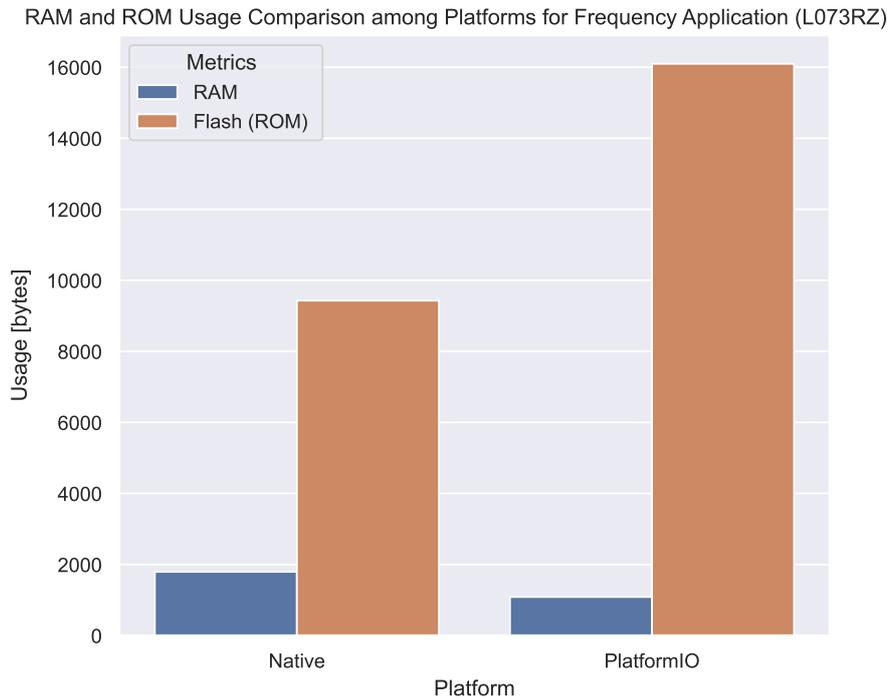

**Figure 2.** RAM and ROM Usage Comparison Among Platforms for Frequency App

**The Frequency App Issue with RIOT**

Unfortunately, RIOT does not easily allow for the modification of the processor frequency. The absence of a low-level interface for frequency management makes conducting this test challenging. To achieve this, delving into the RIOT source code would be necessary, which was not feasible within the scope of this internship due to time constraints. However, research has been conducted revealing that researchers are working on integrating DVFS (Dynamic Voltage and Frequency Scaling) into RIOT OS [18]. At least one of these researchers appears to be an active developer for RIOT and has presented this feature as desirable for the future of this OS [19]. DVFS dynamically manages the processor's frequency and voltage based on workload. This optimizes power consumption by adjusting these parameters in real time, providing better energy efficiency than manual frequency changes.

## 3. Classifier

The most notable observation in Figure 3 is the doubling of RAM usage compared to the previous tool. However, in previous experiments, we noted that PlatformIO favoured ROM usage while consuming less RAM. It is important to contextualize this increase. Nonetheless, the trend of RIOT OS to have higher RAM usage seems to be confirmed in this context. Regarding ROM memory, we observe a 23% increase on the RIOT OS side compared to the PlatformIO tool. This difference aligns with the continuity of previous observations, although it is more significant this time. These findings highlight trends that are consistent with previous experiments.

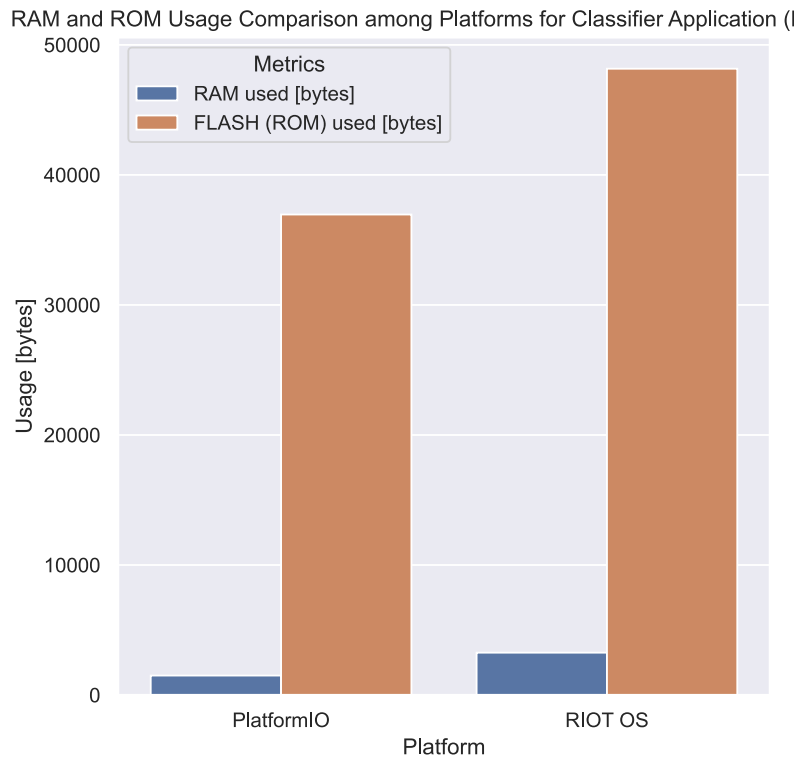

**Figure 3.** Comparison of RAM and ROM Usage Across Platforms for the Classifier App

As expected, the memory cost for the same application is higher with an OS compared to the Arduino Framework and considerably higher than a native approach. However, this impact appears moderate and does not generate particular difficulties for future software implementations on programmable boards.

## VI.B Energy Use Analysis

The measurements were conducted by pairing the target development board with a current measurement board. Both boards are connected to a computer, enabling both the transfer of programs and debugging and the collection of current measurements.

### 1. Blinky

During this experimentation, we conducted a series of tests on the RIOT, Arduino Framework, and STM32CubeIDE environments. We chose the basic Blinky app to evaluate different aspects of the tools. Our primary objective was to measure the average power consumption and the minimum, maximum, and quartile values. This provided us with an overview of the energy behaviour of each tool, enabling us to make an objective compari-

son.Figure 4 shows the analysis of three 100-second experiments for each tool, representing approximately 150 cycles of LED blinking per experiment, ensuring reliable measurements.

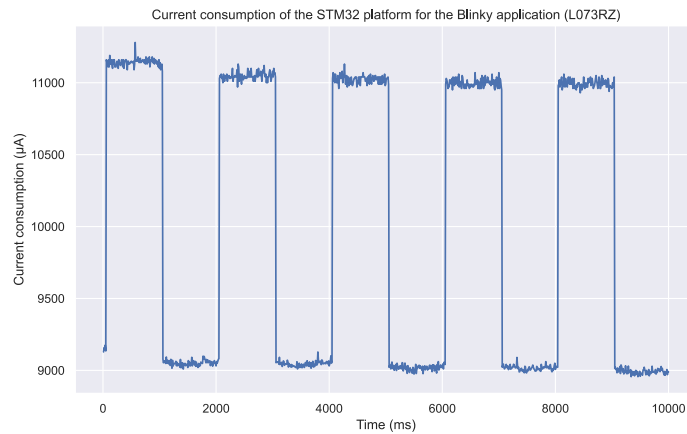

**Figure 4.** Current Consumption of the STM32 Platform for the Blinky Application (Zoomed in on 10 seconds)

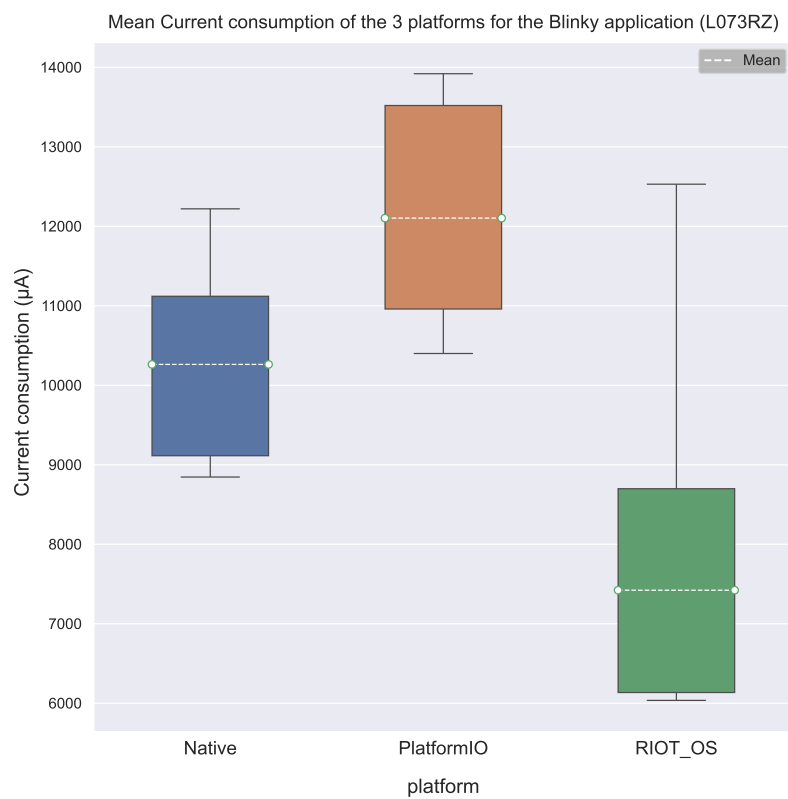

**Figure 5.** Average Current Consumption of 3 Platforms for the Blinky App

Figure 5 illustrates the energy consumption based on the platform. Unsurprisingly, using the Arduino Framework results in higher energy consumption for operation in an equivalent configuration. Surprisingly, RIOT OS consumes approximately 40% less energy than the native program. It is worth noting that there is a slightly higher standard deviation and significantly larger peaks in maximum consumption.

## 2. Frequency

This test was conducted across the Arduino Framework and STM32CubeIDE environments. The goal was to assess the average power consumption for different frequency ranges based on the tool used. The LED briefly turns on between each frequency, as depicted in Figure 6, as we observe consumption spikes.

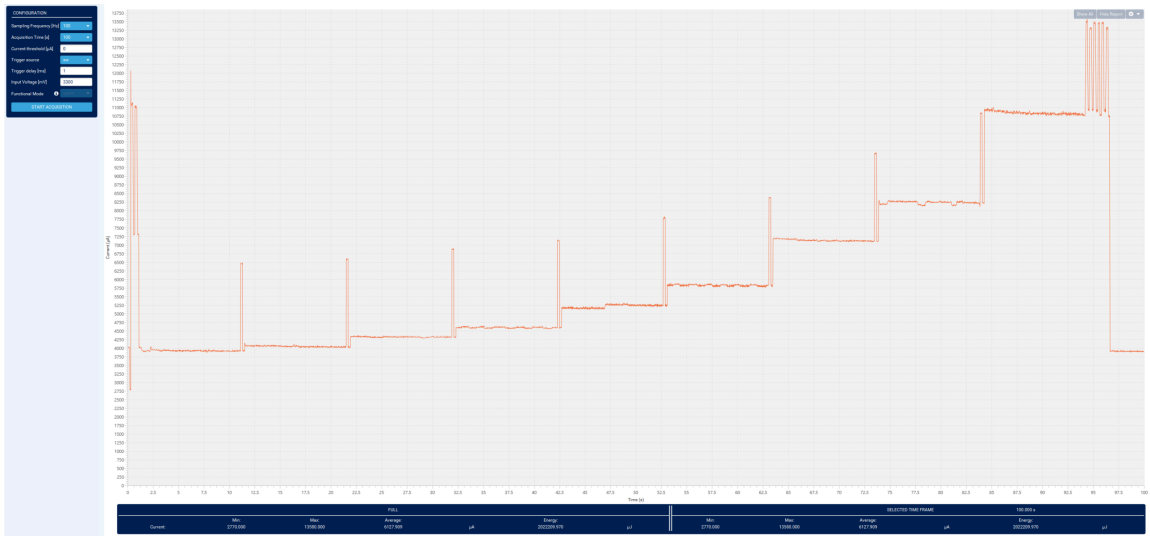

**Figure 6.** Analysis of the Power Consumption of the Frequency Program (using the Arduino Framework) with STM32CubeMonitor-Power Software

Figure 7 illustrates the relationship between the microcontroller frequency and the average current consumed. Here, we find the classic approximation of microcontroller power consumption: $P \propto f \times V^2$ [20]. The curve reveals an extremely low standard deviation across all data. However, it is essential to note that more significant variability occurs in higher frequency ranges, suggesting more pronounced fluctuation levels than lower frequencies. It is worth noting that despite the conclusions drawn from this observation, no automated analysis was conducted using Python code, limiting the possibility of obtaining a more comprehensive and objective evaluation of this data. Integrating such an approach would have allowed for more rigorous results.

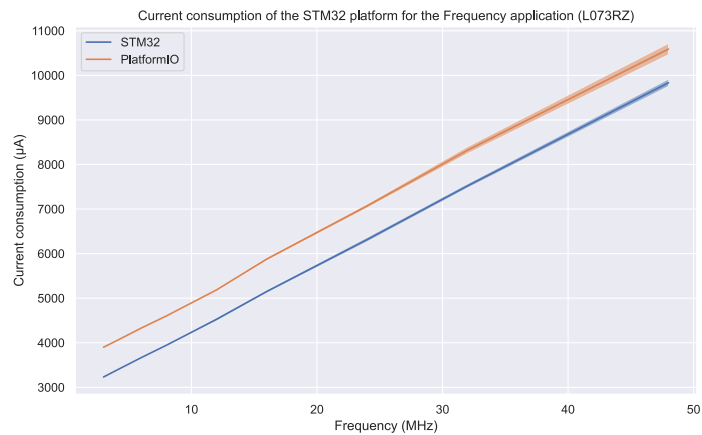

**Figure 7.** Evolution of Average Current Based on Microcontroller Frequency

The main observation lies in the consistency of the gap between the two consumption curves. Although the behaviour evolves similarly for both approaches, it is notable that PlatformIO maintains a higher consumption than the STM32 program. This observation indicates that PlatformIO generates a more power-hungry behaviour, ranging from approximately 700 to 800 µA, regardless of the chosen frequency.

## 3. Classifier

This test was conducted in the Arduino Framework and RIOT environments. The aim was to assess the average consumption for each tool while running this program. We performed five experiments, each lasting 100 seconds. The consumption measurement was taken over the entire execution cycle, and we isolated the classifier's training and testing phases. One can observe a measurement in Figure 8. With RIOT OS, we clearly notice the learning and inference phases.

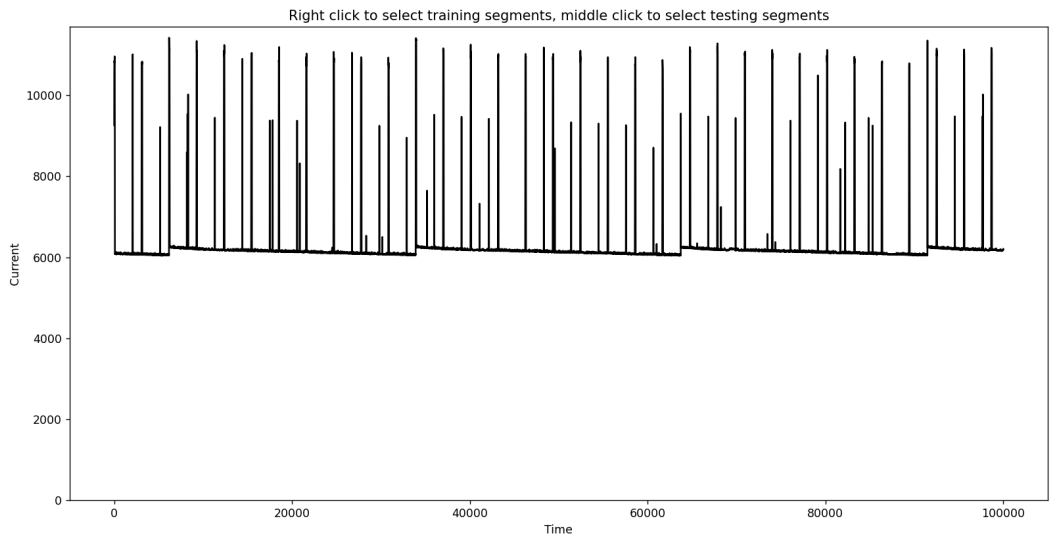

**Figure 8.** Current consumption over time with the Classifier app (RIOT)

To obtain reliable data and isolate the differ, a script was used to visualize consumption graphs by delineating each phase in time, defined by a start and an end (Figure 9). This method exposes some measurement imprecision, but we export the start and end parameters of the windows so that they can be replayed and verified if needed. Zooming in on parts of the figure makes it easier to notice the peaks caused by the learning and inference phases due to their small size. The periods between the red and green bars correspond to the learning phase, while those between the blue and yellow bars correspond to the inference phase.

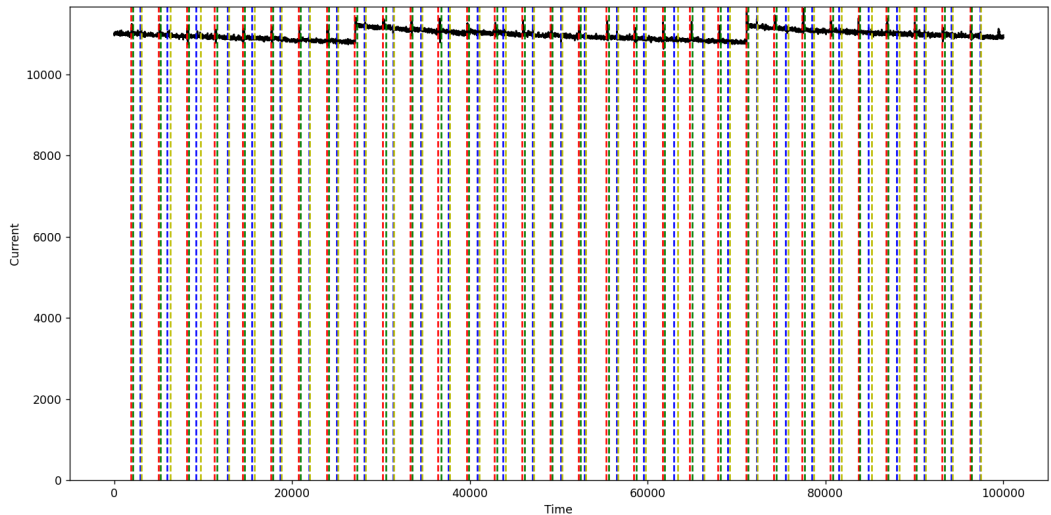

**Figure 9.** Current consumption over time with the Classifier app (Arduino Framework), with delimited periods

**The Classifier Issue with STM32CubeIDE**

The lack of high-level abstraction makes reproducing this program on STM32CubeIDE challenging. Its implementation would require a substantial time investment, and the generated code would likely need to be optimized. Figure 10 allows us to compare the overall average current consumption with the task performed, namely learning and inference.

Surprisingly, RIOT OS consumes 40% less energy than PlatformIO in this context. In the case of RIOT OS, significant standard deviations are observed, much higher than with PlatformIO, as highlighted in Figure 8 with consumption peaks. These standard deviations increase even more during the inference phases, with the highest consumption peaks. Despite this, the average values significantly differ between each phase.

On the other hand, no significant difference in consumption is observed between the different phases when using PlatformIO. Indeed, the averages for the learning and inference phases are pretty close, particularly considering the variation introduced by the standard deviation. Another surprising observation Regarding RIOT OS is that inference is less energy-consuming than learning.

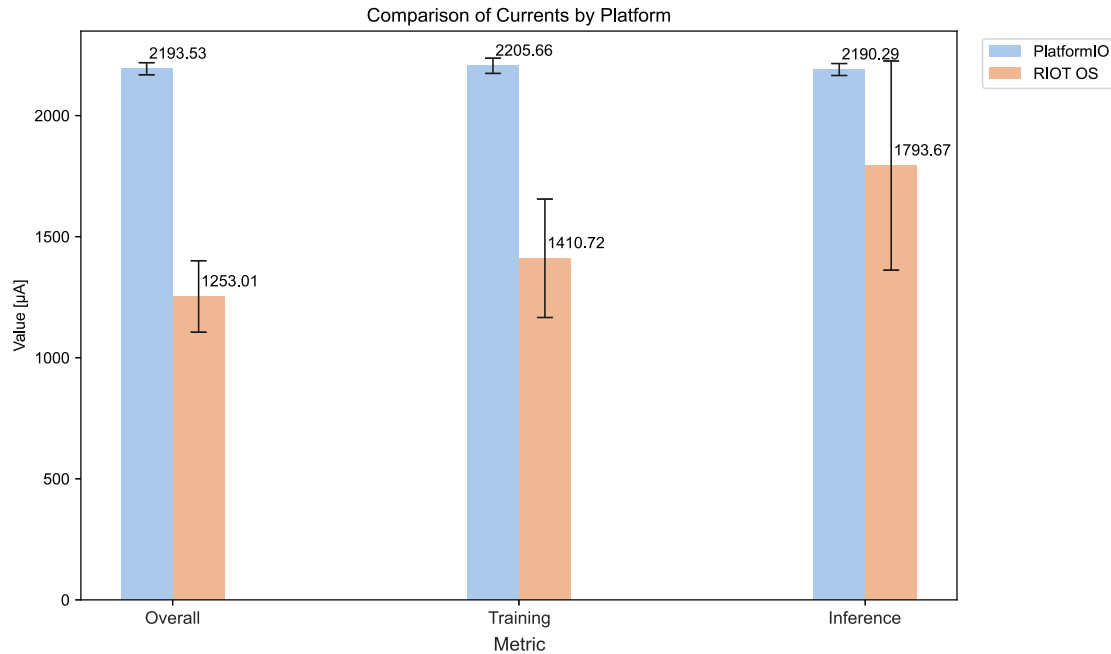

**Figure 10.** Evolution of the overall average current and according to the performed task

RIOT stands out as the clear winner regarding performance and energy consumption. It consumes less in all measured aspects. This difference is explained by RIOT OS being optimized and automatically turning off anything unnecessary. Even though we tried to turn off unused features in STM32CubeIDE, some components may have remained active and might not have been essential.

RIOT presents a significant advantage for development: it facilitates the creation of energy-efficient software. Creating a low-power program is more feasible with this OS than the other two solutions. In the previous section, we observed a peculiar behaviour: the higher consumption during inference, which is explained by using a simplistic dataset in training the perceptron. In the case of real applications, we would need a more complex neural network than a simple isolated neuron, and consumption would likely be higher (although this still needs to be verified by measurements).

### VI.C Performance Analysis of the Classifier

Figure 11 presents the metrics associated with training our ML model with the perceptron. Interestingly, the two curves are nearly superimposed because we applied the same random seed in both cases. This similarity implies that the program functions similarly in both environments, allowing for comparable and consistent energy consumption values between the experiments.

This training and resulting inferences were performed on a subset of 100 rows from the IRIS dataset. We can observe a decrease in the 'Recall Evolution' metric while the other metrics progress upwards. During inference, this relatively simple dataset achieves a 100% success rate. This observation suggests satisfactory performance of the ML model in both environments, providing a basis for consistent energy consumption comparisons between the two experiments (RIOT and Framework).

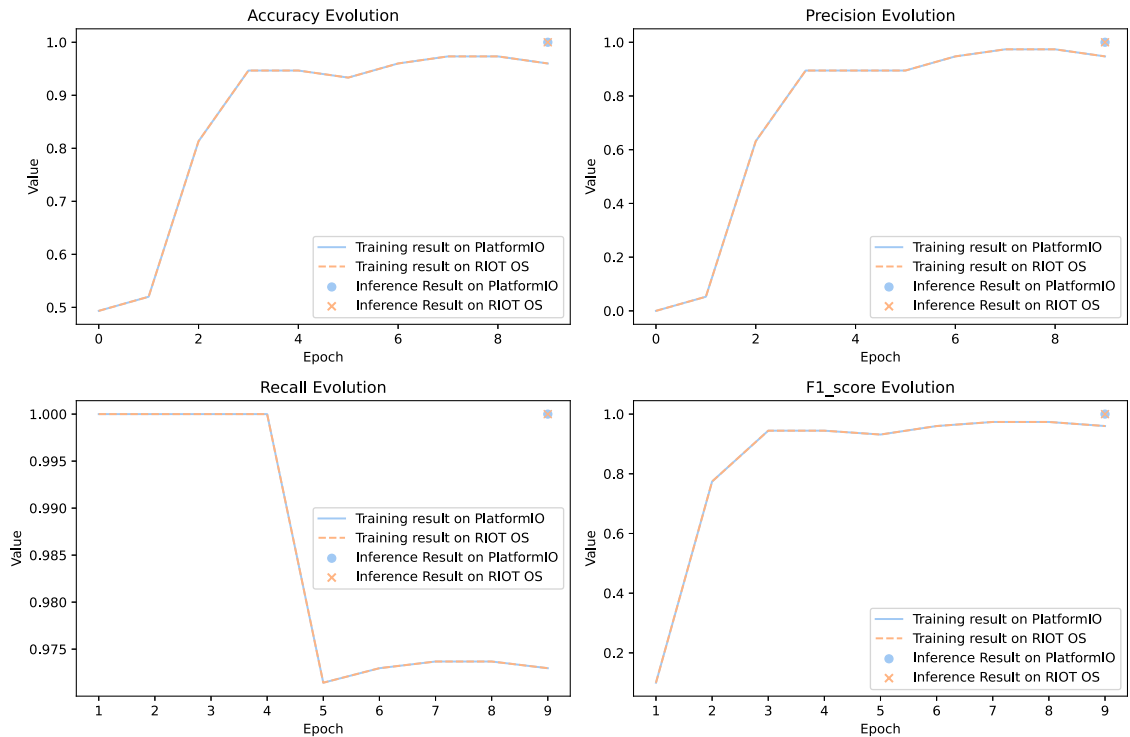

**Figure 11.** Perceptron metrics evolution on Iris Flower dataset

## VII   CONCLUSION

In this study, we tested bare metal (STM32 HAL), middleware (Arduino Framework, PlatformIO), and RTOS (RIOT) development environments in terms of on-device learning performances. RIOT stands out as a good compromise choice for designing complex systems like intelligent sensors, enabling the deployment of small-scale training and inference solutions for tiny ML models. Indeed, the RTOS solution consumes more memory than the native solution but appears less energy-consuming than the intermediate solution, facilitating prototyping. To delve deeper, we might compare it with other OS and use more powerful boards to run more elaborate models and move datasets outside the memory. Indeed, the perceptron is ridiculously small due to dataset sizes, and the tasks do not correspond to the use cases of cyber-physical systems. Unfortunately, critical APIs like DVFS are missing, which would allow for finer control of hardware and energy expenditure.